\documentclass[a4paper, 10pt, conference]{IEEEtran}
\usepackage{amsmath}
\usepackage{amsfonts}
\usepackage{graphicx}
\usepackage{hyperref}
\usepackage{geometry}
\IEEEoverridecommandlockouts                              % This command is only needed if 
\title{\LARGE \bf
Advancing H\&E-to-IHC Stain Translation in Breast Cancer: A Multi-Magnification and Attention-Based Approach
}

\author{Linhao Qu$^{*}$$^{1}$, Chengsheng Zhang$^{*}$$^{2}$, Guihui Li$^{*}$$^{2}$, Haiyong Zheng$^{2}$, Chen Peng$^{\dag}$$^{3}$ and Wei He$^{\dag}$$^{4}$% <-this % stops a space
\thanks{* Linhao Qu, Chengsheng Zhang and Guihui Li contribute equally to this work.}% <-this % stops a space
\thanks{$\dag$ Wei He and Chen Peng are the co-corresponding authors.}% <-this % stops a space
\thanks{$^{1}$ Linhao Qu is with Fudan University, Shanghai 200032, China. {\tt\small lhqu20@fudan.edu.cn}}%
\thanks{$^{2}$ Chengsheng Zhang, Guihui Li and Haiyong Zheng are with College of Electronic Engineering, Ocean University of China, Qingdao 266404, China. {\tt\small guihuilee@163.com, zcs@stu.ouc.edu.cn, zhenghaiyong@ouc.edu.cn}}%
\thanks{$^{3}$ Chen Peng is with Research Center for Frontier Fundamental Studies, Zhejiang Lab, Hangzhou 311121, China. {\tt\small pc1527918@163.com}}%
\thanks{$^{4}$ Wei He is with Institutes of Biomedical Sciences, Fudan University, Shanghai 200032, China. {\tt\small 16301050918@fudan.edu.cn}}%
}

\begin{document}
%new
\clearpage % 开始新的一页
\newgeometry{
  top=94pt, % 修改上页边距
  bottom=68pt, % 修改下页边距
  left=40pt, % 保持左页边距不变
  right=40pt % 保持右页边距不变
}
%new

\maketitle
\thispagestyle{empty}
\pagestyle{empty}

%%%%%%%%%%%%%%%%%%%%%%%%%%%%%%%%%%%%%%%%%%%%%%%%%%%%%%%%%%%%%%%%%%%%%%%%%%%%%%%%
\begin{abstract}
Breast cancer presents a significant healthcare challenge globally, demanding precise diagnostics and effective treatment strategies, where histopathological examination of Hematoxylin and Eosin (H\&E) stained tissue sections plays a central role. Despite its importance, evaluating specific biomarkers like Human Epidermal Growth Factor Receptor 2 (HER2) for personalized treatment remains constrained by the resource-intensive nature of Immunohistochemistry (IHC). Recent strides in deep learning, particularly in image-to-image translation, offer promise in synthesizing IHC-HER2 slides from H\&E stained slides. However, existing methodologies encounter challenges, including managing multiple magnifications in pathology images and insufficient focus on crucial information during translation. To address these issues, we propose a novel model integrating attention mechanisms and multi-magnification information processing. Our model employs a multi-magnification processing strategy to extract and utilize information from various magnifications within pathology images, facilitating robust image translation. Additionally, an attention module within the generative network prioritizes critical information for image distribution translation while minimizing less pertinent details. Rigorous testing on a publicly available breast cancer dataset demonstrates superior performance compared to existing methods, establishing our model as a state-of-the-art solution in advancing pathology image translation from H\&E to IHC staining.
\end{abstract}

%%%%%%%%%%%%%%%%%%%%%%%%%%%%%%%%%%%%%%%%%%%%%%%%%%%%%%%%%%%%%%%%%%%%%%%%%%%%%%%%
\section{INTRODUCTION}
Breast cancer remains a principal cause of cancer-related mortality among women globally. Reducing mortality necessitates precise diagnostics and effective treatment strategies. Histopathological examination, central to the diagnostic regimen, typically involves analyzing tissue sections stained with Hematoxylin and Eosin (H\&E), serving as the initial screening approach \cite{qu2024rethinking,qu2023boosting}. Pathologists diagnose breast cancer by examining these H\&E-stained slides directly under a microscope or by reviewing digital whole slide images (WSIs) generated from whole slide scanners \cite{14,qu2022bi}.

Once breast cancer is diagnosed, evaluating specific biomarkers like Human Epidermal Growth Factor Receptor 2 (HER2) is essential for personalized treatment plans. HER2 overexpression correlates with breast cancer's aggressive behavior, necessitating its precise detection to guide therapy choices. Currently, HER2 levels are primarily assessed through Immunohistochemistry (IHC), which uses antibodies to visualize HER2 proteins in tumor cells. Although informative, IHC demands costly equipment and specialized skills, limiting its use in resource-constrained settings \cite{1}. In contrast, H\&E staining is more affordable, and the digital processing of these stains presents opportunities for technological advances. Developing an algorithm that can automatically transform H\&E stained WSIs to mimic IHC staining could significantly reduce costs and reliance on specialized human resources.

Recent advancements in deep learning, particularly in image-to-image translation technologies, offer promising solutions for synthesizing IHC-HER2 slides from H\&E stained slides. This approach involves training deep neural networks, such as Generative Adversarial Networks (GANs) \cite{2}, to learn mappings between the source domain (H\&E slides) and the target domain (IHC slides), capturing the translation from H\&E to IHC staining characteristics. In this context, generative adversarial training utilizes both aligned pairwise data and unpaired data. Techniques like Pix2pix \cite{3} and Pyramid Pix2pix \cite{4} facilitate translation through a pixel-to-pixel approach using registered H\&E-IHC paired datasets. However, achieving perfect pixel-level alignment between H\&E and IHC images is challenging. Typically, slides are prepared from two adjacent tissue layers, each stained separately with H\&E and IHC, which introduces discrepancies due to variations in cell morphology, staining-induced damage (e.g., tissue tearing), and scanning artifacts, thereby limiting these methods' effectiveness. Conversely, methods such as CycleGAN \cite{5}, CUT \cite{6}, and ASP \cite{7} operate effectively with unpaired data. CUT \cite{6} and ASP \cite{7} enhance generative adversarial training by integrating contrastive learning, thereby improving their performance in transforming pathological images from H\&E to IHC under less constrained conditions.

Despite promising advancements, current image-to-image translation methods in pathology face two primary challenges: \textbf{1) Handling of Multiple Magnifications}: Pathological images, unlike natural images, require analysis across various magnifications due to their complex and layered features. For example, specific cellular arrangements and morphologies are discernible at a higher magnification (40x), whereas the broader microenvironment and tumor boundaries are more apparent at a lower magnification (5x). Capturing and integrating information from these various magnifications is crucial for accurate distribution translation, a task that current methodologies struggle to accomplish efficiently. \textbf{2) Inadequate Focus on Crucial Information}: the effectiveness of feature extraction in image translation processes varies across
\clearpage % 开始新的一页
\restoregeometry % 恢复原始页边距
\noindent different image regions. It is essential for networks, particularly those employing contrastive learning and generative adversarial techniques, to prioritize information that significantly impacts the translation of image distributions while allocating less attention to less impactful details. 

To address the challenges in pathological image translation from H\&E to IHC staining, we propose a novel model that incorporates attention mechanisms and multi-magnification information processing. To tackle the first issue, our model employs a multi-magnification processing strategy that extracts and utilizes information from different magnifications within the pathological images during the translation training. This approach enables robust image translation across various magnifications. For the second challenge, we have integrated an attention module within the generative network. This module prioritizes information critical for image distribution translation during feature extraction from pathological images, while minimizing the influence of less pertinent details.

The effectiveness of our algorithm was rigorously tested on a publicly available breast cancer H\&E to IHC dataset using both subjective assessments and objective metrics. The evaluation results demonstrate that our model achieves superior performance compared to existing methods, establishing it as a state-of-the-art solution in the field.

\section{Related Work}
Currently, in the translation of pathological images from H\&E to IHC staining, prominent approaches involve generative adversarial training using aligned pair-wise data and unpaired data. Methods include Pix2pix \cite{3} and Pyramid Pix2pix \cite{4}, which utilize aligned pair-wise data. Pix2pix utilizes the source domain image as a conditional input to the generator and computes L1 loss \cite{8} directly between the generated fake IHC-stained image and the real IHC image (ground truth). Pyramid Pix2pix employs a low-pass filter to smooth images, downsamples to reduce resolution and eliminate redundant pixels, and calculates L1 loss on high-dimensional magnification features. However, achieving pixel-level perfect alignment between H\&E and IHC images is unattainable, thereby constraining their performance.

CycleGAN \cite{5}, CUT \cite{6}, and ASP \cite{7} represent methods that do not rely on perfectly aligned image pairs. CycleGAN introduces cycle consistency loss, performing forward and backward image translations concurrently. CUT integrates contrastive learning into unsupervised non-paired image translation problems by enforcing the quality of generated images through contrastive loss computed between the input image to the generator and the generated fake image. ASP, an extension of CUT, introduces adaptive supervised contrastive loss, which reduces the impact of inconsistency during training by weakening the loss derived from image pairs with low correspondence.

While existing methods have shown promising performance, they encounter two primary challenges: 1) inadequate processing of information from multiple magnifications in pathological images, 
and 2) difficulty in effectively prioritizing and extracting critical information for distribution translation during image feature extraction. In contrast, we propose a novel pathological image translation model for converting H\&E to IHC staining, leveraging attention mechanisms and multi-magnification information processing. To address the first challenge, we integrate an attention module into the generative network. This module focuses on crucial information for image distribution translation during feature extraction from pathological images, while minimizing the impact of less important details. To tackle the second challenge, we introduce a multi-magnification processing strategy that extracts information from different magnifications within pathological images for translation training. This enables the model to perform robust image translations across various magnifications.

\section{Method}

\begin{figure*}[!h]
  \includegraphics[width=\textwidth]{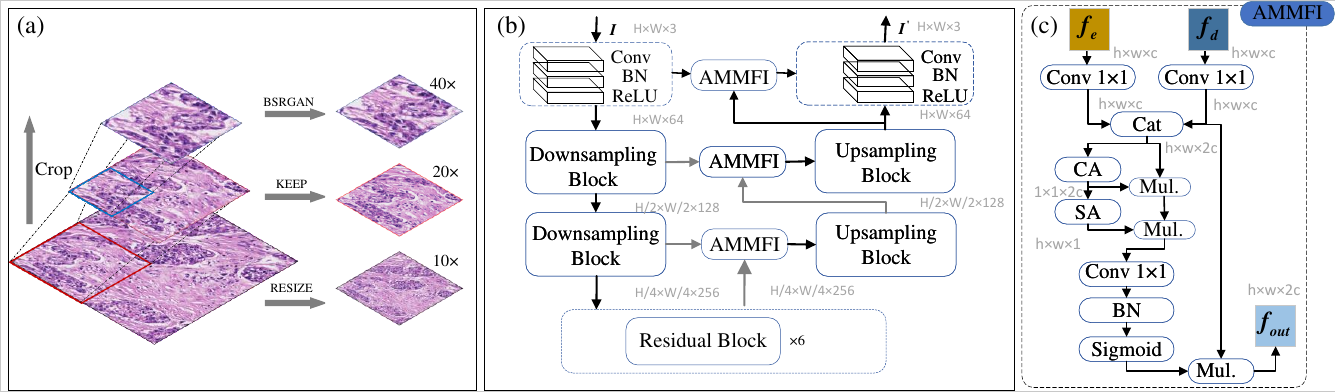}
  \caption{(a) Our proposed multi-magnification pathology image processing strategy. (b) The overall structure of the proposed generator. (c) AMMFI refers to the attention-based multi-magnification feature interaction module. Conv denotes a 3x3 convolutional layer, while BN signifies a batch normalization layer. BSRGAN is identified as a super-resolution model. Mul. indicates an element-wise multiplication operation, and Cat represents concatenation. CA is an abbreviation for channel attention, and SA stands for spatial attention.}
  \label{fig:1}
\end{figure*}

This paper is dedicated to H\&E-to-IHC stain image translation, aiming to learn mapping functions $G(x \rightarrow y)$ and/or its inverse $F(y \rightarrow x)$ from one staining modality (H\&E) to (IHC), ensuring that the transformed images $P_x$, after being processed by $F$, closely resemble the target domain $P_y$, while also maximizing the similarity between $P_y$, after undergoing $G$ translation, and the original $P_x$. To achieve this goal, we design an attention-based multi-magnification image translation model, depicted in Fig. \ref{fig:1}, built around the foundational principles of Generative Adversarial Networks (GANs) \cite{3}. Central to our method is the novel introduction of the multi-magnification processing strategy and the attention module integrated in the generator network. The discriminator in our method is referred to PatchGAN \cite{3}.

\subsection{Proposed Generator Overview}
As is shown in Fig. \ref{fig:1} (b), given a H\&E image $I \in \mathbb{R}^{H \times W \times 3}$ at any magnification level, it is first passed through a layer of $3 \times 3$ convolutional kernels for initial processing, extracting shallow features $f_s \in \mathbb{R}^{H \times W \times 64}$. These shallow features are then fed into a two-level encoding module, gradually refining to generate more abstract and expressive deep features $f_e \in \mathbb{R}^{H/4 \times W/4 \times 256}$. Subsequently, $f_e$ undergoes further enhancement through a series of carefully designed residual blocks (ResNet Blocks), enhancing the feature's expressive power and discriminative ability. Building upon this, we propose an attention-based multi-magnification feature interaction module (AMMFI), which integrates visual features from the decoding level with corresponding features from the encoding level, achieving precise localization and enhancement of core pathological details at different magnification levels. Ultimately, these focused and refined features flow through a recursive decoding process, progressively upsampled and recombined, to output a high-fidelity feature representation $ I^{\prime}  \in \mathbb{R}^{H \times W \times 3} $ that perfectly matches the original image space dimensions, thus completing the task of high-fidelity image translation from one staining modality to another.

\subsection{Multi-magnification Processing Strategy}

We present a multi-magnification processing strategy for translating pathology images. We develop a dataset comprising images at various magnifications to facilitate deep learning models in learning pathology image features across different scales. This processing strategy includes several key steps: First, we apply downsampling techniques to the original high-resolution images to simplify details while retaining the original magnification and emphasizing macroscopic structural contours. Next, we use random cropping and zooming techniques to increase the size of selected local regions to twice and four times their original size, respectively. This step is aimed at enhancing the model's ability to recognize local fine features such as cell morphology and tissue texture, thereby providing microscopic insights. Finally, all processed images are adjusted to a uniform size through an existing pre-trained super-resolution model \cite{10} to ensure the standardization and efficiency of model training.

% We propose a multi-magnification processing strategy for pathology image translation. To do this, we construct a multi-magnification pathology image dataset to enhance the learning of deep learning models on multi-magnification features of pathology images. Specifically, the multi-magnification pathology image processing consists of the following steps: firstly, utilizing downsampling techniques to moderately compress the original high-resolution images, aiming to maintain the original magnification while simplifying details, highlighting macroscopic structural contours. Subsequently, employing random cropping and zooming techniques to enlarge selected local regions to twice and four times the original size, aiming to deeply explore and enhance local fine features, such as cell morphology and tissue texture, injecting microscopic insights into the model. Finally, all images processed through different methods are adjusted to a uniform size through super-resolution model \cite{10} to ensure the standardization and efficiency of model training.

\subsection{Attention-based Multi-magnification Feature Interaction Module}

We introduce an attention-based multi-magnification feature interaction module (AMMFI) aimed at optimizing the feature processing pipeline and enhancing the model's spatial focus capabilities through advanced gating mechanism \cite{15} and attention mechanism \cite{16}. This module is especially effective in pathology image translation tasks, where it significantly enhances the accuracy and utility of the results. Our design integrates a gating mechanism with spatial and channel attention mechanisms. It processes concatenated features from various encoder layers and initial decoder stages, applying refined attention to both channels and spatial dimensions to precisely enhance feature saliency.

The unique aspect of our module is the use of gating signals, customized into a grid pattern that reflects key spatial information of the image. This allows for tailored adjustments across different image regions. Furthermore, skip connection strategies enable these gating signals to aggregate features from multiple magnifications, improving the resolution of the processing signal and thereby excelling in complex image scenarios.

The module autonomously identifies and localizes critical pathological regions within an image, effectively filtering out irrelevant background noise through attention mechanisms. This ensures a highly accurate translation of essential pathological biomarkers. In essence, the module dynamically focuses on multi-magnification features, improving both the precision and interpretability of the pathological image analysis.

The pipeline of our proposed AMMFI is defined as follows:

\begin{equation}
f^{\prime} = Cat((W_{f_{e}}^T f_{e}^{i} + b_{f_{e}}),  (W_{f_{d}}^T f_{d}^{j} + b_{f_{d}})),
\end{equation}

\begin{equation}
f^{\prime \prime} = M_c(f^{{{\prime}}}) \otimes f^{{{\prime}}},
\end{equation}

\begin{equation}
f_{out}^{j}=\sigma\left(W_{a}\left(M_s\left( M_c(f^{{{\prime}}})\right) \otimes f^{\prime \prime} \right)+b\right) \otimes f_{e}^{i},
\end{equation}
where $f_{e}^{i}$ represents the feature output from the i-th layer of the encoder, $f_{d}^{j}$ represents the feature output from the j-th layer of the decoder corresponding to the encoding layer. $ W_{f_{e}}^T$, $ W_{f_{d}}^T$, and $W_a$ represent weight parameter matrices, $\sigma$ represents the sigmoid activation function, $b_{f_{e}}$, $b_{f_{d}}$ and $b$ represent bias matrices. $M_c$ and $M_s$ respectively represent channel attention and spatial attention. 
$Cat(\cdot)$ stands for concatenation, and $\otimes$ represents element-wise multiplication.

\subsection{Training Loss Functions}
To maintain visual fidelity, structural color consistency, and semantic relevance of content in the image translation, we employ a series of loss functions to guide model training. Specifically, we combine the following types of loss functions:

\textbf{Adversarial Loss.} This loss function originates from the framework of Generative Adversarial Networks (GANs), which introduces a discriminator network to encourage generated images to closely match the distribution of real images \cite{1}. The adversarial loss encourages generated images to be visually indistinguishable from real images, thereby enhancing their naturalness and realism.

\textbf{Adaptive Weighted Supervised Contrastive Loss.} This loss is referred to as Adaptive Supervised PatchNCE Loss \cite{8}, which computes by first selecting patches from real IHC-stained images and generated fake IHC-stained images, then utilizes an adaptive weighting scheme to calculate the contrastive loss. It is defined as:

\begin{equation}
w_{t}(v,v^{+}) = (1 - g(t/T)) \times 1.0 + g(t/T) \times h(v \cdot v^{+})),
\end{equation}

\begin{equation}
L_{ASP} = w_{t}(v,v^{+}) L_{infoNCE}(v,v^{+},v^{-}),
\end{equation}
where $v$ is the feature vector of the query patch, $v^{+}$ is the feature vector of the positive patch, $v^{-}$ is the feature vector of the negative patch, $t$ represents the current iteration number during model training, and $T$ denotes the total number of iterations. $g(\cdot)$ is the weight scheduling function, it is used to adjust the weights at different stages of training. $h(\cdot)$  is the similarity weight function, it is employed to address the issue of a non-strict pairing of image pairs, which can alleviate the negative impact of inconsistent targets on training. $L_{infoNCE}(v,v^{+},v^{-})$ loss is derived from \cite{7}.

\textbf{Gaussian Loss.} The Gaussian Loss is an improvement upon the loss function used in PixPix \cite{5}. In Pix2Pix, only the $L1$ loss is computed at the image level between the generated fake images and the real images. To mitigate this limitation, the Gaussian Loss employs specific Gaussian kernels to perform multiple convolutions and downsampling on the images, computing the $L1$ loss across different feature layers, and finally aggregating them with weighted summation. It is defined as:
\begin{equation}
L_{GP} = \sum_{i}\lambda_{i} S_{i},
\end{equation}
where $i$ represents the feature layer, $S_{i}$ denotes the $L1$ loss computed for the i-th layer, and $\lambda_{i}$ is the weighting coefficient.

Finally, the overall loss function of the generator is defined as:

\begin{equation}
\begin{aligned}
L &= \lambda_{adv} L_{adv} + \quad \lambda_{PatchNCE} L_{PatchNCE} + \quad \lambda_{ASP} L_{ASP} \\ +
   &\quad \lambda_{GP} L_{GP},
\end{aligned}
\end{equation}
where $L_{adv}$ represents the traditional adversarial loss function, $L_{PatchNCE}$ \cite{7} represents the contrastive loss function, $ L_{ASP}$ represents the adaptive weighted supervised contrastive loss function, $ L_{GP}$ represents the Gaussian loss function, and $\lambda$ represents the weights of each loss function.

\begin{table*}[t!]
\centering
\caption{Quantitative evaluations of our method and the competitors. The bold parts represent the optimal metrics.}
\resizebox{\textwidth}{!}{%
\begin{tabular}{ccccccccccc}
\hline
Method        & SSIM↑           & PHV\_layer1↓    & PHV\_layer2↓    & PHV\_layer3↓    & PHV\_layer4↓    & PHV\_avg↓       & FID↓          & KID↓         & LPIPS↓          & PSNR↑            \\ \hline
Pix2pix \cite{3}       & 0.1559          & 0.5516          & 0.5070          & 0.3253          & 0.8511          & 0.5588          & 137.3         & 82.9         & 0.4722          & 14.2550          \\
CycleGAN \cite{5}     & 0.1914          & 0.5633          & 0.6346          & 0.4695          & 0.8871          & 0.6386          & 240.3         & 311.1        & 0.4598          & 13.5389          \\
CUT \cite{6}           & 0.1810          & 0.5321          & 0.4826          & 0.3060          & 0.8323          & 0.5383          & 66.8          & 19.0         & 0.4542          & 14.3022          \\
PyramidP2P \cite{4}   & 0.2078          & 0.4787          & 0.4524          & 0.3313          & 0.8423          & 0.5362          & 104.0         & 61.8         & 0.4493          & 14.2033          \\
ASP \cite{7}           & 0.2004          & 0.4534          & 0.4150          & 0.2665          & 0.8174          & 0.4881          & \textbf{51.4} & 12.4         & 0.4546          & 14.1371          \\ \hline
\textbf{Ours} & \textbf{0.2153} & \textbf{0.4482} & \textbf{0.4094} & \textbf{0.2639} & \textbf{0.8157} & \textbf{0.4843} & \textbf{51.4} & \textbf{8.6} & \textbf{0.4432} & \textbf{14.3711} \\ \hline
\end{tabular}}
\label{tab1}
\end{table*}

\begin{table*}[t!]
\centering
\caption{Results of the ablation study. The bold parts represent the optimal metrics. }
\resizebox{\textwidth}{!}{%
\begin{tabular}{ccccccccccc}
\hline
Method                  & SSIM↑           & PHV\_layer1↓    & PHV\_layer2↓    & PHV\_layer3↓    & PHV\_layer4↓    & PHV\_avg↓       & FID↓          & KID↓         & LPIPS↓          & PSNR↑            \\ \hline
Our Baseline            & 0.2004          & 0.4534          & 0.4150          & 0.2665          & 0.8174          & 0.4881          & \textbf{51.4} & 12.4         & 0.4546          & 14.1371          \\
\textbf{w/. multi-manification} & 0.1980          & 0.4494          & 0.4119          & 0.2654          & 0.8165          & 0.4858          & 60.6          & 29.7         & 0.4465          & 14.3046          \\
\textbf{w/. Attention}  & \textbf{0.2187} & 0.4527          & 0.4148          & 0.2867          & \textbf{0.8183} & 0.4918          & 61.6          & 23.0         & 0.4478          & 14.2963          \\ \hline
\textbf{Ours}           & 0.2153          & \textbf{0.4482} & \textbf{0.4094} & \textbf{0.2639} & 0.8157          & \textbf{0.4843} & \textbf{51.4} & \textbf{8.6} & \textbf{0.4432} & \textbf{14.3711} \\ \hline
\end{tabular}}
\label{tab2}
\end{table*}

\begin{figure*}[t!]
  \centering
  \includegraphics[width=0.9\textwidth]{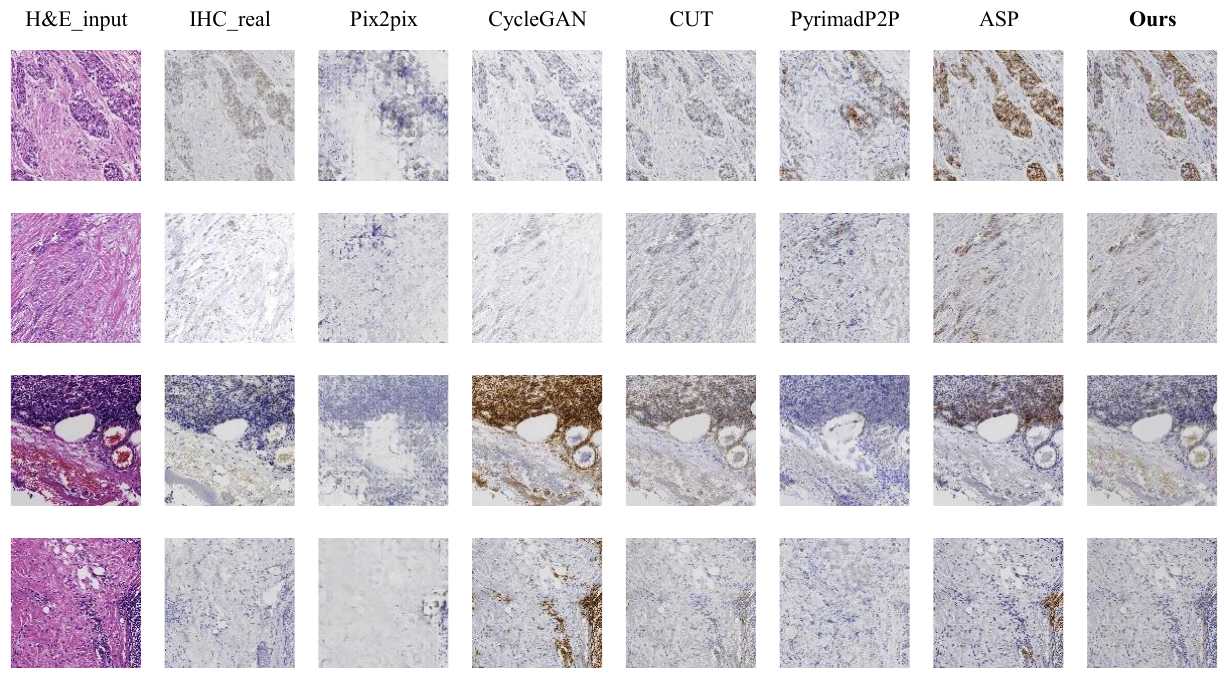}
  \caption{Visual results of our method and the competitors. "H\&E\_input" and "IHC\_real" respectively represent the input image stained with H\&E and the real image stained with IHC.}
  \label{fig:2}
\end{figure*}

\begin{figure*}[h!]
  \centering
  \includegraphics[width=0.85\textwidth]{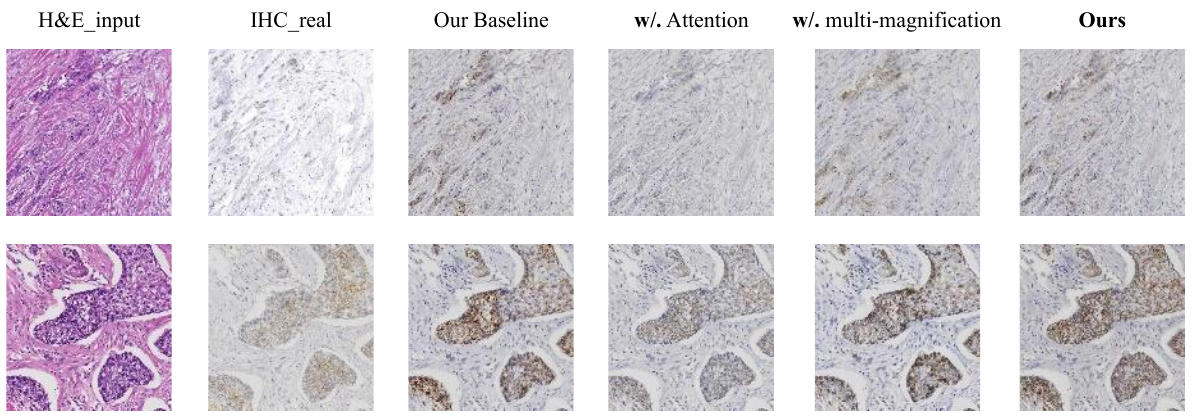}
  \caption{Visual results of the ablation study. "H\&E\_input" and "IHC\_real" respectively represent the input image stained with H\&E and the real image stained with IHC. "Our Baseline" means that our proposed attention module and multi-magnification strategy are not used. "w/. Attention" and "w/. multi-magnification" mean that only our proposed attention module or multi-magnification strategy is used respectively. "Ours" means that both of the strategies are used.}
  \label{fig:3}
\end{figure*}

\section{Experiments}
\subsection{Datasets}
We utilized the publicly available breast cancer dataset MIST-HER2 (Multi-IHC Stain Translation-HER2 \cite{7}) to comprehensively evaluate the performance of our algorithm. This dataset comprises 4642 pairs of training images and 1000 pairs of test images extracted from 64 Whole Slide Images (WSIs). The data consists of paired images, each including an H\&E image and its adjacent slice stained with IHC-HER2. All patches are sized at 1024×1024 pixels.

\subsection{Implementation Details}
We employed ResNet-6Blocks with the proposed attention module as the generator and a 5-layer PatchGAN \cite{3} as the discriminator. During training without multi-magnification pathology image processing strategy, images were randomly cropped to $512\times512$ size with a batch size of 1. For training with the multi-magnification pathology image processing strategy, images were processed at multiple resolutions according to our proposed training paradigm, also with a batch size of 1. We utilized the Adam optimizer with a linear decay scheduler, starting with an initial learning rate of $2\times10^{-4}$. The coefficients for the loss functions were set as follows: $\lambda_{adv}=1$, $\lambda_{PatchNCE}=10$, $\lambda_{ASP}=10$, $\lambda_{GP}=10$. All experiments were conducted using the Python programming language, PyTorch 1.11 deep learning framework, and executed on a computing platform equipped with an NVIDIA GeForce RTX 3090 GPU.

\subsection{Evaluation Metrics}
We utilized both paired and unpaired evaluation metrics to compare these methods. For comparing generated and real images, we employed standard metrics such as SSIM (Structural Similarity Index Measure \cite{ssim}), PSNR (Peak Signal-to-Noise Ratio \cite{psnr}), and LPIPS (Learned Perceptual Image Patch Similarity \cite{lpips}), as well as PHV (Perceptual Hash Value \cite{13}) . For assessing unpairedness, we utilized FID (Fréchet Inception Distance \cite{fid}) and KID (Kernel Inception Distance \cite{kid}).

\subsection{Competitors}
We compared the five latest deep learning-based methods for H\&E-IHC translation: Pix2pix \cite{3} and Pyramid Pix2pix \cite{4} trained on aligned pairs, and CycleGAN \cite{5}, CUT \cite{6}, and ASP \cite{7} trained on unpaired data.

\subsection{Qualitative Evaluations}
Fig. \ref{fig:2} intuitively illustrates the generated images of our method compared to the comparative methods. It can be observed that our method, utilizing attention mechanisms and a multi-magnification pathology image processing strategy, produces IHC images with more distinct pathological features, closely resembling the real stained IHC images.

\subsection{Quantitative Evaluations}
Table \ref{tab1} presents the objective metrics comparing our method with the comparative methods. It can be observed that our method outperforms all comparative methods across all metrics, achieving the best performance, which aligns with the visual results.

\subsection{Ablation Study}
We conducted ablation experiments on the proposed multi-magnification pathology image processing strategy and attention module, and the objective metric results are shown in Table \ref{tab2}. "Our Baseline" means that our proposed attention module and multi-magnification strategy are not used. "w/. Attention" and "w/. multi-magnification" mean that only our proposed attention module or multi-magnification strategy is used respectively. "Ours" means that both of the strategies are used. It can be observed that incorporating either the multi-magnification pyramid processing strategy or the attention module alone leads to significant improvements compared to the baseline. Furthermore, when both are combined, all metrics show substantial enhancements, confirming the effectiveness of our proposed strategies. We further visualize the visual effects of generating IHC images with each module separately in Fig. \ref{fig:3}. Combining Fig. \ref{fig:2} and Fig. \ref{fig:3}, our model consistently produces images closer to the gold standard in terms of both overall performance and image details.

\section{CONCLUSIONS}
This paper proposes a novel deep learning model for synthesizing IHC-stained slides from H\&E stained slides to address the challenges in evaluating specific biomarkers like HER2 in breast cancer diagnostics. By integrating multi-magnification information processing and attention mechanisms, our model demonstrates superior performance compared to existing methods in pathology image translation. Rigorous testing on publicly available breast cancer datasets validates the effectiveness of our approach, establishing it as a state-of-the-art solution for advancing pathology image translation from H\&E to IHC staining. This contribution holds significant promise for enhancing the precision and efficiency of breast cancer diagnostics, potentially leading to improved treatment strategies and patient outcomes in the future.

\bibliographystyle{IEEEtran}
\bibliography{ref}

\end{document}